\documentclass[review,number,sort&compress]{elsarticle}

\usepackage{caption}

\usepackage{subfig} 
\usepackage{rotating}

\usepackage{lineno}  

\usepackage{graphicx}  

\usepackage{xspace}  
\usepackage{color}
\usepackage{colortbl}
\usepackage{rotating}
\usepackage{multirow}
\usepackage{bigints}
\usepackage{amsmath}
\usepackage{ifthen} 
\usepackage[
    plainpages=false,
    bookmarks         = true,
    bookmarksnumbered = true,
    pdfpagemode       = None,
    pdfstartview      = FitH,
    pdfpagelayout     = SinglePage,
    colorlinks        = true,
    linkcolor         = blue,
    pagecolor         = black,
    urlcolor          = blue,
    pdfborder         = {0 0 0}
    ]{hyperref}%
\newboolean{uprightparticles}
\setboolean{uprightparticles}{false} 
\usepackage{amssymb}
\usepackage{amsfonts}
\usepackage{upgreek}
\usepackage{rotating}
\usepackage{multirow}
\usepackage{mciteplus}
\usepackage{booktabs}
\definecolor{miriam}{rgb}{0.65, 0.04, 0.37}
 \def\PJ      {\ensuremath{J}\xspace}

 \def\Pi      {\ensuremath{i}\xspace}

\newcommand{\CLs}{\ensuremath{\textrm{CL}_{\textrm{s}}}\xspace}

\def\Ppsi {\ensuremath{\psi}\xspace}
\def\jpsi     {{\ensuremath{{\PJ\mskip -3mu/\mskip -2mu\Ppsi\mskip 2mu}}}\xspace}
\def\Jpsi     {{\ensuremath{\jpsi}}\xspace}

\def\Bbar    {\kern 0.18em\overline{\kern -0.18em B}{}\xspace}

\newcommand{\Bs}{\ensuremath{B^0_s}\xspace}

\mathchardef\PLambda="7103

\newcommand{\BsJpsiKK}{\ensuremath{B^0_s\to \jpsi K^+K^-}\xspace}

\newcommand{\BsJpsiPhi}{\ensuremath{B^0_s\to \jpsi \phi}\xspace}

\def\Kbar  {\kern 0.2em\overline{\kern -0.2em K}{}\xspace}

\newcommand{\tev}{\ifthenelse{\boolean{inbibliography}}{\ensuremath{~T\kern -0.05em eV}\xspace}{\ensuremath{\mathrm{\,Te\kern -0.1em V}}}\xspace}
\newcommand{\gev}{\ensuremath{\mathrm{\,Ge\kern -0.1em V}}\xspace}
\newcommand{\mev}{\ensuremath{\mathrm{\,Me\kern -0.1em V}}\xspace}
\newcommand{\kev}{\ensuremath{\mathrm{\,ke\kern -0.1em V}}\xspace}
\newcommand{\ev}{\ensuremath{\mathrm{\,e\kern -0.1em V}}\xspace}
\newcommand{\gevc}{\ensuremath{{\mathrm{\,Ge\kern -0.1em V\!/}c}}\xspace}
\newcommand{\mevc}{\ensuremath{{\mathrm{\,Me\kern -0.1em V\!/}c}}\xspace}
\newcommand{\gevcc}{\ensuremath{{\mathrm{\,Ge\kern -0.1em V\!/}c^2}}\xspace}
\newcommand{\gevgevcccc}{\ensuremath{{\mathrm{\,Ge\kern -0.1em V^2\!/}c^4}}\xspace}
\newcommand{\mevcc}{\ensuremath{{\mathrm{\,Me\kern -0.1em V\!/}c^2}}\xspace}

\newcommand{\Ah}{\ensuremath{A^0}\xspace}

\newcommand{\pdf}{\ensuremath{{\rm PDF}}\xspace}

\newcommand{\swave}{{\rm S--wave}\xspace}
\newcommand{\pwave}{{\rm P--wave}\xspace}
\newcommand{\dwave}{{\rm D--wave}\xspace}

\newcommand{\tabref}[1]{Table~\ref{#1}}

\newcommand{\secref}[1]{Sect.~\ref{#1}}

\journal{None}

\begin{document}

\begin{frontmatter}
\title{Ipanema-$\beta$ : tools and examples for HEP analysis on GPU}
\author[a,b]{D. Mart{\'{\i}}nez Santos}
\author[c] {P. \'Alvarez Cartelle}
\author[b]{M. Borsato}
\author[b] {V. G. Chobanova}
\author[b]{ J. Garc{\'{\i}}a Pardi\~nas}
\author[b]{M. Lucio Mart{\'{\i}}nez}
\author[b] {M. Ramos Pernas}
\address[a]{Axencia Galega de Innovaci\'{o}n. Conseller\'{\i}a de Industria. Xunta de Calicia, Spain }
\address[b]{Universidade de Santiago de Compostela, Santiago de Compostela , Spain}
\address[c]{Imperial College London, London, United Kingdom}
\begin{abstract}
\noindent 

\end{abstract}



%



\end{frontmatter}



%


\section*{Acknowledgements}
 
\noindent We would like to thank Cedric Potterat and the rest of our colleagues at the Federal University of Rio de Janeiro, for introducing us to
the world of GPUs and for showing us the large gain in computing speed that such devices can achieve in a likelihood fit.
We are very grateful for the funding provided by EPLANET, which made this exchange possible. We would also like to thank Axencia Galega
de Innovacion (GAIN), XuntaGAL and the European Research Council (ERC-StG-639068) for their support.

\cleardoublepage
\newpage

\section{Introduction}
\label{sec:intro}

We present here a set of examples, classes and tools which can be used for statistical analysis in Graphics Processing Units (GPU). This includes binned and unbinned maximum
likelihood fits, pseudo-experiment generation, convolutions, Markov Chain Monte Carlo method implementations, and limit setting techniques.
Our idea is that these examples can be used either as starting points by anyone interested in the subject, or as a framework to develop given analyses.
Hence, we include cases with different levels of complexity, starting by showing how to use and combine existing packages, down to more realistic examples that
required of us the development of generic classes.
Our code is based on pyCUDA~\cite{pycuda}, which is then combined with scientific packages for analysis such as iMinuit~\cite{iminuit, minuit} or pyMultinest~\cite{pymultinest}.
In \secref{sec:others}, we list the packages that we use, as well as other approaches we tried or that could  be of interest for the reader. Sections \ref{sec:example1}-\ref{sec:toyMC}
describe the provided examples. 
Section \ref{sec:faq} is meant to collect some frequently asked questions.

\clearpage
\newpage
\section{Software}
\label{sec:others}
The \texttt{Ipanema} package is available on \texttt{git}:
\begin{center}
\texttt{git clone https://gitlab.cern.ch/bsm-fleet/Ipanema.git .}
\end{center}
It contains the following folders:
\begin{center}
  \texttt{urania cuda polibio examples }
\end{center}
\noindent and a simple \texttt{setup.csh} script.
The software we use depends on the following:
\begin{itemize}
\item Cuda. Tested with 7.5 and 8.0. 
\item \texttt{Python2} (we used 2.7.11).
\item Standard \texttt{Python} packages for scientific usage, such as scipy, numpy, matplotlib. We got them all through an installation of Anaconda \cite{Anaconda}. We tested both against Anaconda2 and Anaconda4.
\item pyCuda \cite{pycuda} (not in Anaconda). In addition, we made a modified version of \texttt{curand.py}, (\texttt{toyrand.py}) where the generation of an array of poisson numbers uses a separate
poisson mean per generated value. We include \texttt{toyrand.py} in the repository.
\item \texttt{reikna} \cite{reikna} (not in Anaconda).
\item Default (for likelihood fits): \texttt{iminuit} \cite{iminuit} (not in Anaconda).  In order to use analytical gradients, one needs to get the version from the \texttt{git} master branch. If not, then the
  standard \texttt{1.2} is fine.
\item Optional, multidimensional parameter sampling: Multinest \cite{Multinest}, pymultinest \cite{pymultinest} (not in Anaconda).
\item Optional, symbolic programing: \texttt{sympy} (included in Anaconda\footnote{The released examples are for \texttt{sympy 0.7.6.1} as well as \texttt{sympy 1.0}, older versions of \texttt{sympy} may need some small tweaking of 
the expression manipulation.}), Mathematica (optional, can be called by \texttt{urania} modules to substitute some of the utilities of \texttt{sympy}).
\end{itemize}

There are several interesting packages on the market that we don't use here, but that the reader may be interested in:
\begin{itemize}
\item GooFit. This is the software that was shown to us in Rio. It mimics the structure of RooFit, but for Cuda. We don't use
it because we preferred to work mostly from \texttt{Python}, and because at the time we knew about GooFit, there were still functionalitites
missing that we needed. It can be used with a related software, Hydra.
\item numba.accelerate: It contains an alternative to pyCUDA, which can upload \texttt{Python} fucntions to a GPU using decorators. It was the base
  of our first attempts to make fitting programs via \texttt{Python}+CUDA. But after some time we moved to pyCUDA instead. We gave it a short try to
  combine both, but they didn't seem compatible.
\item Faddeeva's function from \href{https://github.com/aoeftiger/faddeevas/blob/master/cernlib\_cuda/wofz.cu}{\texttt{GitHub}}. It is not used in the current
  released examples, but we are using it for more realistic versions of the $B_{s}^{0}\to\Jpsi KK$ fit, for instance.

To test the examples we provide, we use three graphics card models: GeForce 980 GTX Ti, GeForce 1080 GTX and Tesla M2090. Their specifications are listed in~\ref{sec:Hardware}. 
\end{itemize}

\clearpage
\newpage
\section{Example 1: Fit to a signal peak on top of an exponential background}
\label{sec:example1}

File: \texttt{examples/sbfit.py}\\
In this example, a signal peak is fitted on top of an exponential background, using an unbinned maximum likelihood fit.
The signal probability density function (PDF) is implemented in a file called \texttt{psIpatia.cu} as a  device function, \texttt{\_\_device\_\_ double log\_apIpatia}.
A \texttt{\_\_device\_\_} function is only accessible by the GPU. The signal PDF is then
processed inside \texttt{sbfit.py} by pyCUDA's \texttt{SourceModule}. Since \texttt{SourceModule} uses strings, we
read \texttt{psIpatia.cu} as:
\begin{center}
\texttt{mod = SourceModule(file("psIpatia.cu","r").read())}.
\end{center}
An alternative way is to define a string which contains the CUDA code, as done in the standard pyCUDA examples.

The signal PDF is later called by a global function, \texttt{\_\_global\_\_ void Ipatia}, over an array of data. 
Global functions are visible both for the GPU and for the host (e.g. a CPU). They must be of type \texttt{void}, meaning that any output has to be passed as an argument.
In this case, the output (\texttt{int\_gpu}) is an array with the signal PDF value of each data point. The function is then called in \texttt{Python} as:
\begin{flushleft}
\texttt{sig\_pdf(bins\_gpu, int\_gpu, np.float64(mu), np.float64(sigma), 
\newline np.float64(l), np.float64(beta), np.float64(a), np.float64(n), 
\newline np.float64(a2), np.float64(n2), block = (1000,1,1), 
\newline grid = (len(mydat)/1000,1))}.
\end{flushleft}
Note that each parameter has to be passed to the GPU using the variable type specified in the CUDA code. In this case, all parameters are of type \texttt{double}, which is passed as \texttt{np.float64(parameter)}.
The last two arguments of the fuction define the block and grid size in the GPU. The maximum block size depends on the computing capability of the graphics card and is smaller or equal than 1024.
The limitations for the grid size are much looser. In this case we are making blocks of 1000 events each, and then the grid as large as needed so that it covers all the events, i.e,
$Nevts=block[x]\times grid[x]$. The $y$ and $z$ components of the block/grid are not used.

The minimization is done by \texttt{Minuit}, via the \texttt{iminuit} package. \texttt{Minuit} minimizes a given cost function, \texttt{FCN}. In most of the usages in high-energy physics, the FCN is $-2\log L$ (or just $-\log L$)
and a typical definition for fit errors correspond to a change of one unit in $-2\log L$.
Inside the \texttt{FCN} the signal and background PDF, 
\[ 
\begin{array}{lll}
  \!\!\!\texttt{bkg\_gpu}\!\!\!\!\!&=&\!\!\!\!\!\texttt{pycuda.cumath.exp(k*data\_array\_gpu)},\\
\end{array}
 \] 
and afterwards the likelihood for each data point are calculated in parallel by the GPU,
\[ 
\begin{array}{lll}
  \!\!\!\texttt{LL\_gpu}\!\!\!\!\!&=&\!\!\!\!\!\texttt{pycuda.cumath.log(fs*invint\_s*sig\_gpu+fb*invint\_b*bkg\_gpu)}.\\
\end{array}
 \] 
Here, \texttt{invint\_s} and \texttt{invint\_b} are the inverses of the signal and background PDF integrals, \texttt{integral\_ipa} and \texttt{integral\_exp}, respectively. 
The calculation of \texttt{integral\_ipa} is performed numerically by summing up the \texttt{int\_gpu} entries,
\begin{center}
 \texttt{integral\_ipa = np.sum((int\_gpu).get())*DM}.
\end{center}
The background PDF integral is determined analytically,
\begin{flushleft}
    \texttt{if k!= 0 :} \\
    \hspace{0.5cm}  \texttt{integral\_exp = (np.exp(k*Mmax)-np.exp(k*Mmin))*1./k} \\
    \texttt{else :}  \\
    \hspace{0.5cm} \texttt{integral\_exp = (Mmax - Mmin)}.
\end{flushleft}

\begin{figure}
 \center
 \includegraphics[width=0.75\textwidth]{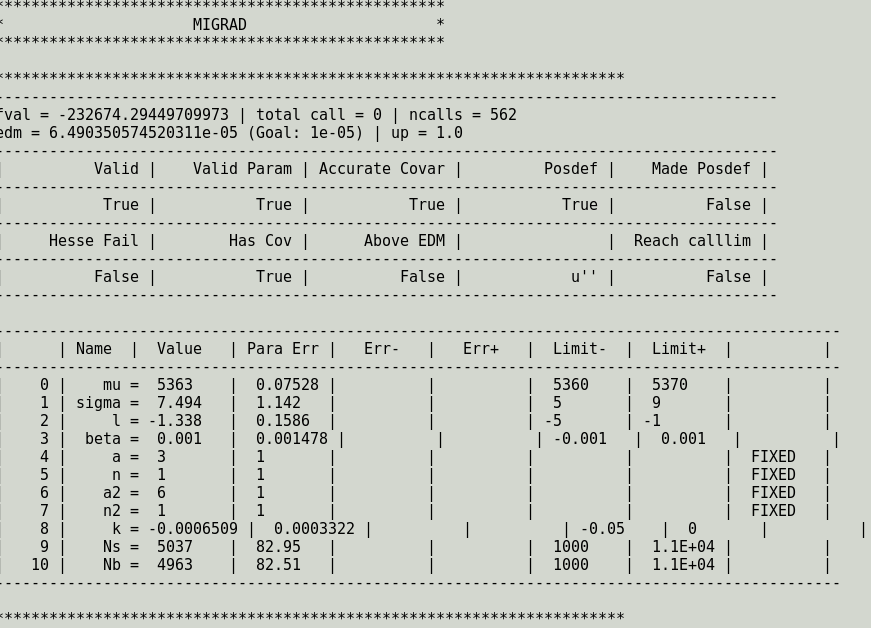}
 \put(5,150){\bf (a)}
 
 \includegraphics[width=0.75\textwidth]{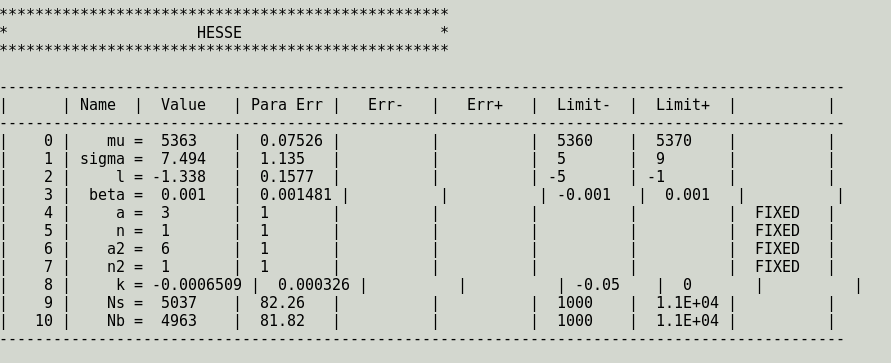}
 \put(5,100){\bf(b)}
 
 \includegraphics[width=0.75\textwidth]{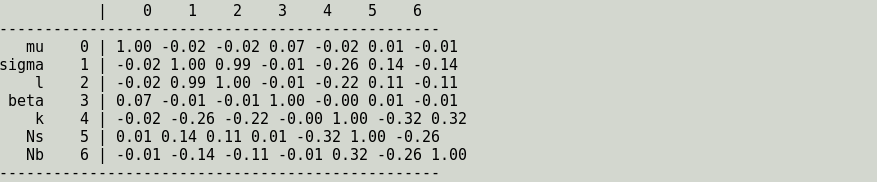}
 \put(5,40){\bf(c)}
 \caption{Fit result of Example 1: (a) \texttt{MIGRAD} output, (b) \texttt{HESSE} output, (c) parameter correlation matrix.}
 \label{fig:Ex1Results}
\end{figure}

This parallelization significantly reduces the computational time compared to an algorithm which loops over the data and calculates each value in sequence. Note also that some operations, 
such as the calculation of an exponential or a logarithmic function, can be performed directly on the \texttt{gpu\_array} from the \texttt{Python} function even though it is parallelized and executed in the GPU, as shown above. In the end, the individual likelihoods are summed 
up to a total likelihood in the CPU\footnote{Depending on the size of the array, in some cases it is faster to pass the entire array to the host and let it do the sum, while above
a given size {\it which we didn't calculate accurately} it becomes better to do the sum in the GPU.}. The sum, including normalization terms, is returned as the output of \texttt{FCN}. 
The results of the \texttt{MIGRAD} and \texttt{HESSE} methods and the corresponding correlation matrix are shown in Fig.~\ref{fig:Ex1Results}.

The example prints in the end the GPU time in seconds. We measured the following timings in different devices (\tabref{tab:timings1}):

\begin{table}[htb!]
\begin{center}
\caption{Timings of Example 1 in different devices (see text for details).}
\label{tab:timings1}
\begin{tabular}{ccc}
 \toprule
 Device & 10k events & 10M events\\
 & \multicolumn{2}{c}{time [s]}\\
\midrule
GeForce 980 GTX Ti & 0.9 & 30.9\\
GeForce 1080 GTX & 1.0 & 24.1\\
Tesla M2090 & 1.1 & 21.5 \\
 \bottomrule
\end{tabular}
\end{center}
\end{table}

\section{Example 2: Fit to the time and angular distribution in bins of the $KK$ mass}
\label{sec:example2}
File: \texttt{examples/phis/fitphis.py}\\
This example bases on the fit used by the LHCb collaboration to determine the CP-violating phase $\phi_{s}$ in the $B_{s}^{0}\to J/\psi K K$ decay channel~\cite{BsJpsiphi}. 
In the model we consider contributions from 
$\phi\to KK$ (P wave), which has three polarization states, and from $f_{0}\to KK$ (S wave). We fit the time dependent angular distributions of $B_{s}^{0}$ and $\bar{B}_{s}^{0}$ 
decays in six $KK$ mass bins. The events are produced with \texttt{EvtGen}~\cite{EvtGen}, 
where equal numbers of $B_{s}^{0}$ and $\bar{B}_{s}^{0}$ are generated according
to a theoretical model, which for simplicity does not include any experimental effects. 

We use a simultaneous fit over a total of 12 categories: six $KK$ mass bins times two $B$ flavours.
Most of the fit parameters are shared between all of the 12 categories. These are: the polarization fractions, the CP-violating phases for each contributing resonance 
and polarization, the physics parameters $\Delta m_{s}$, $\Gamma$, $\Delta \Gamma$. The S wave parameters are individual for each bin, since a priori they differ between the $KK$ mass bins.
The external constant $C_{SP}$ accounts for the $KK$-mass integral in the  interference between the S and the P waves. These factors are imported from a \texttt{Python} list
from the file \texttt{phisParams.py}.

The parameter handling is performed by the classes \texttt{Parameter} or \texttt{Free} defined in \texttt{ModelBricks.py}. In \texttt{Parameter}, a
parameter is initialized, from a parameter name, and an initial value. Additional information can be passed at the initialization, such as, limits, step, type and a flag indicating whether it is a constant (default).
\texttt{Free} inherits from \texttt{Parameter}, with the only difference that the default is \texttt{constant=False}. For example, to fix the longitudinal fraction, we use
\begin{center}
\texttt{fL = Parameter("fL",0.5, limits=(0.3,.6))},
\end{center}
while to set it free in the fit:
\begin{center}
\texttt{fL = Free("fL",0.5, limits=(0.3,.6))}.
\end{center}

The latter would be equivalent to:
\begin{center}
\texttt{fL = Parameter("fL",0.5, limits=(0.3,.6), constant = False)},
\end{center}

\noindent as well as to
\begin{center}
\texttt{fL = Parameter("fL",0.5, limits=(0.3,.6))}\\
\texttt{fL.constant = False}
\end{center}

The categories for the simultaneous fit are created with a custom class, \texttt{Cat}, defined in \texttt{ModelBricks.py}. Instances of \texttt{Cat} store information such as the data array, 
and default values block and grid size to be used for each bin when calculating the likelihood. As \texttt{Python} objects, additional attributes can be incorporated {\it ad-hoc} without
existing in the definition of the \texttt{Cat} class. In this example, we add the bin number, a pointer to the function that calculates the integral for that category
(the integral function is different for \Bs and $\bar{\Bs}$), etc. When initializing an instance of \texttt{Cat}, a GPU array with the size of the data is reserved in order to store the likelihood probability values 
at each point .

A list of parameters and categories is passed when initializing an instance of the model class, \texttt{myModel}, 
\begin{center}
\texttt{manager = myModel(Params, cats)}
\end{center}
defined in \texttt{PhisModel.py}. The class \texttt{myModel} inherits from the \texttt{ParamBox} class
defined in \texttt{ModelBricks.py} and is adapted to the particular problem in this example. For details on \texttt{ParamBox} see \ref{sec:ParamBox}.

In \texttt{myModel}, the PDF and its normalization are calculated for each data point by the function
\begin{center}
\texttt{run\_cat(self,cat,CSP,Fs,fL,fpe,dpa,dpe,dS,G,DG,DM,phis)} 
\end{center}
and then combined to a total $-2\log L$. The corresponding physics distributions are taken from the physics model 
defined in \texttt{cuda\_phis.c}. Once an instance of the model is initialized, a \texttt{Minuit} fit can be created using the class method \texttt{createFit()} of \texttt{ParamBox}. 
The class method \texttt{createFit()} instantiates a \texttt{Minuit} object from \texttt{iminuit} and passes to it central values, ranges and other information of the parameters.
The \texttt{Minuit} instance is stored as \texttt{fit} attribute of \texttt{manager}.

\begin{figure}
 \includegraphics[width=0.9\textwidth]{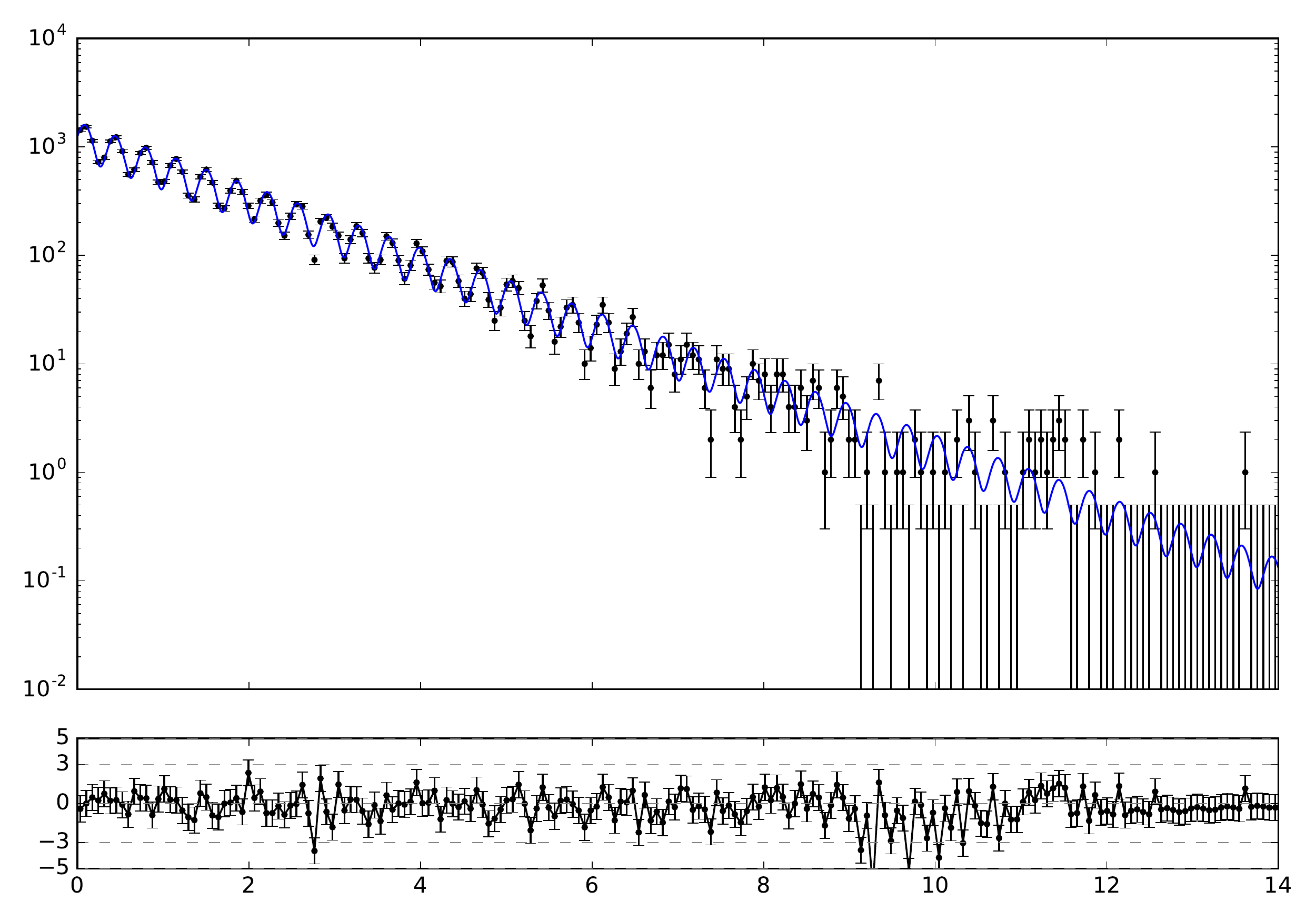}
 \caption{Plot of the fit result of Example 2: Decay time distribution in the fourth $KK$ mass bin, and flavour $\Bs$.}
 \label{fig:ex2}
\end{figure}

\begin{figure}
 \center
 \includegraphics[width=0.75\textwidth]{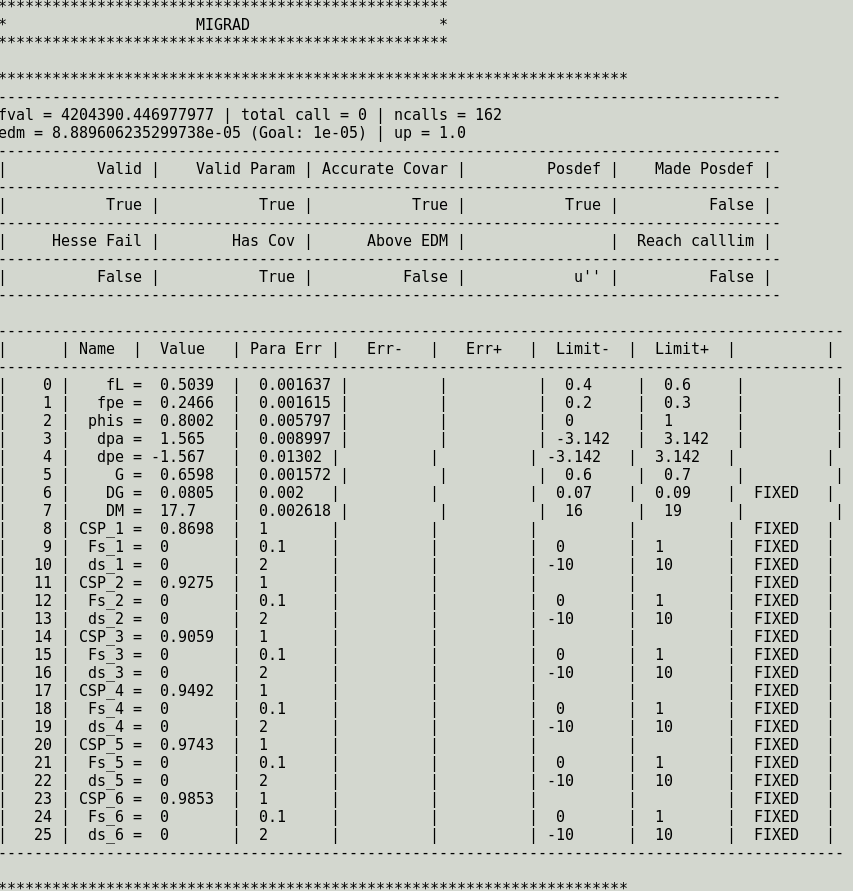}
 \put(5,200){\bf (a)}
 
 \includegraphics[width=0.75\textwidth]{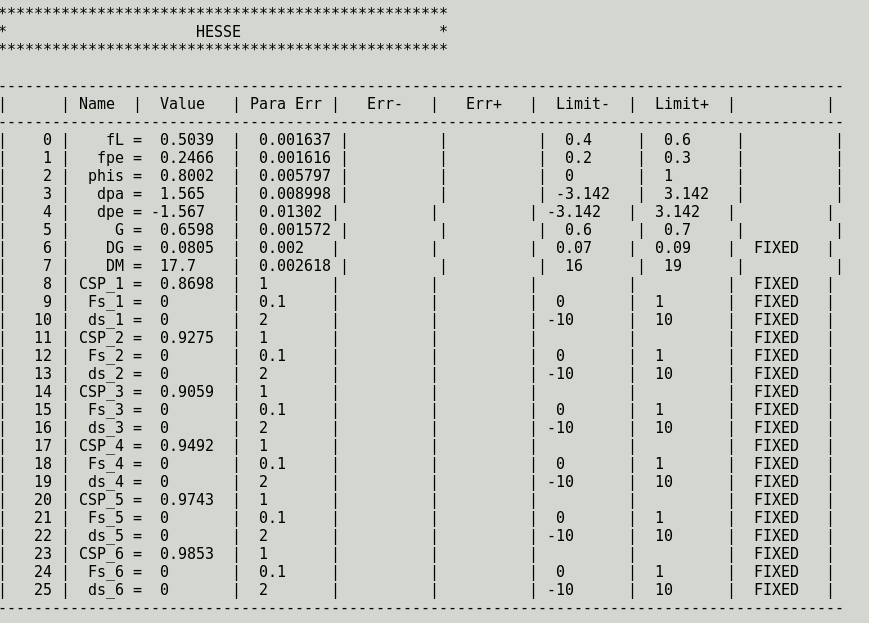}
 \put(5,100){\bf(b)}
 
 \includegraphics[width=0.75\textwidth]{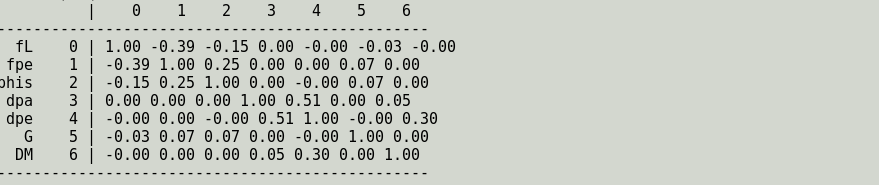}
 \put(5,40){\bf(c)}
 \caption{Fit result of Example 2 for the case of a P wave contribution only: (a) \texttt{MIGRAD} output, (b) \texttt{HESSE} output, (c) parameter correlation matrix.}
 \label{fig:Ex2Results}
\end{figure}

A method \texttt{plotcat(Cat)} creates a plot (via \texttt{matplotlib}) of the result in a given bin including the pulls, as shown in Fig.~\ref{fig:ex2}. The 
fit result for the case when only the P wave contributes to the final state is shown in Fig.~\ref{fig:Ex2Results}. The computation times as tested on different devices
are summarized in Tab.~\ref{tab:timings2}.

\texttt{ParamBox} also includes a method that can launch a scan of the parameter space using \texttt{MultiNest} ,
\begin{center}
\texttt{manager.createMultinest("mnest\_party")}.
\end{center}
\noindent This is done via \texttt{pymultinest}. The string passed to \texttt{createMultinest} is the folder in which \texttt{MultiNest} will store its outputs.
Although \texttt{pymultinest} has a default value for it (``./chains''), here we force the user to set it manually, to avoid reading from an unrelated chain
made on the same folder. Keyword arguments will be sent to \texttt{pymultinest}. In addition, a keyword argument \texttt{reset = True/False} can be used to restart
the sampling if a scan on the same folder was done before. Then, the \texttt{ParamBox} class provides methods to read the  \texttt{MultiNest} output, for example
\begin{center}
\texttt{manager.mnest\_vals()}
\end{center}
\noindent prints out the most probable value. The default priors in \texttt{ParamBox} are flat, thus the printed values should be close to those found by \texttt{Minuit},
unless one of the two algorithms has converged to a local minimum or the parameter space was undersampled. Note that convergence in \texttt{MultiNest} will be way slower than for \texttt{Minuit} (although \texttt{MultiNest}
can offer other advantages).

\begin{table}[htb!]
\begin{center}
\caption{Timings of the Example 2 in different devices (see text for details).}
\label{tab:timings2}
\begin{tabular}{ccc}
\toprule
 Device & 10k events & 10M events\\
 & \multicolumn{2}{c}{time [s]}\\
\midrule
GeForce 980 GTX Ti & 1.9 & 60\\
GeForce 1080 GTX & 1.8 & 50\\
Tesla M2090 & 1.9 & \\
 \bottomrule
\end{tabular}
\end{center}
\end{table}
 
\section{Example 3: Binned fit to $\Upsilon\rightarrow\mu^+\mu^-$ decays in bins of $\eta$ and $p_T$ and search for $\Ah\rightarrow\mu^+\mu^-$ decay}
\label{sec:example3}
\begin{figure}[t]
  \begin{center}
    \subfloat[$\text{Bin}_{p_T} = 0~~~~\text{Bin}_{\eta} = 0$]{
      \includegraphics[scale=0.25]{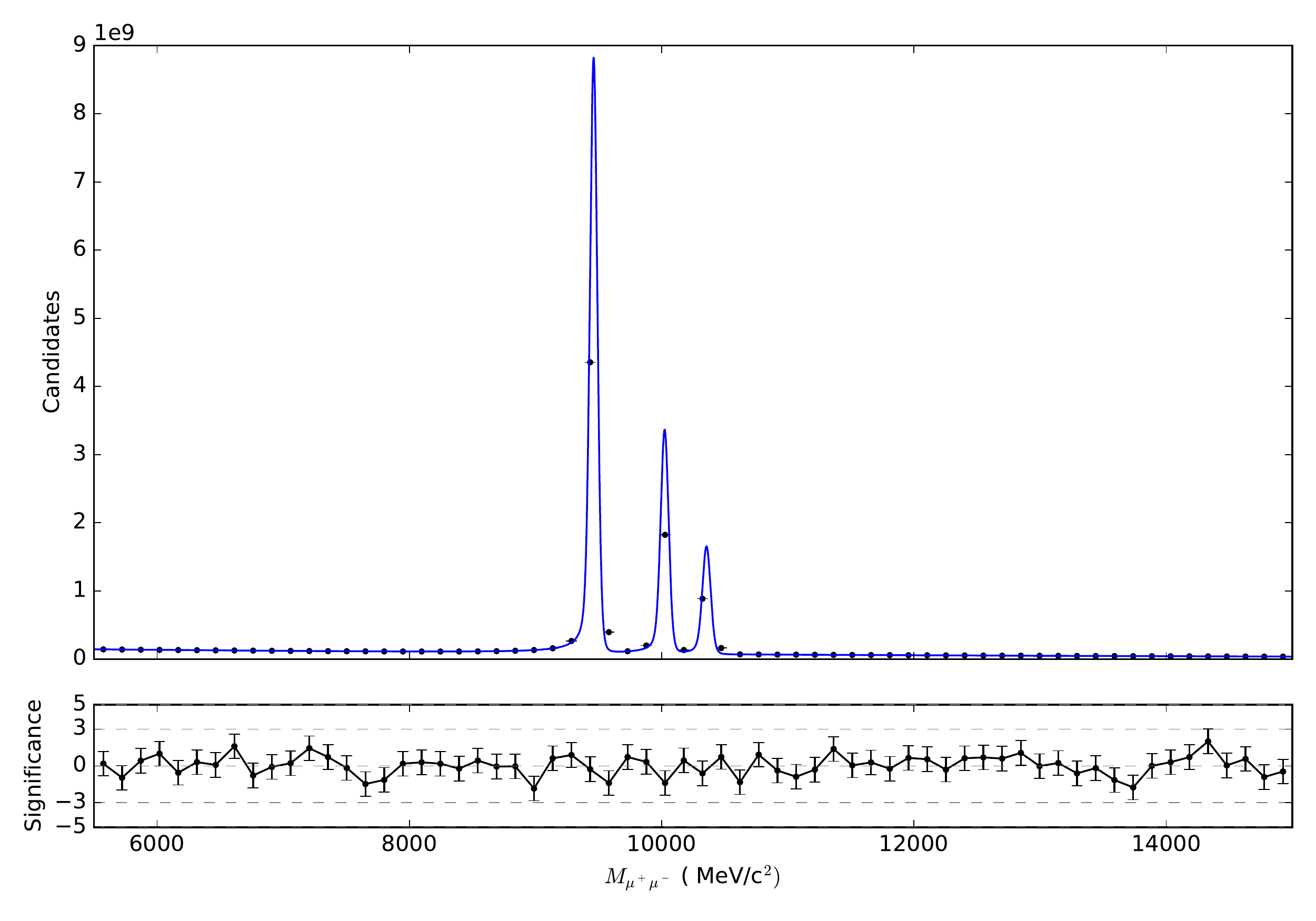}}
    \subfloat[$\text{Bin}_{p_T} = 0~~~~\text{Bin}_{\eta} = 1$]{
      \includegraphics[scale=0.25]{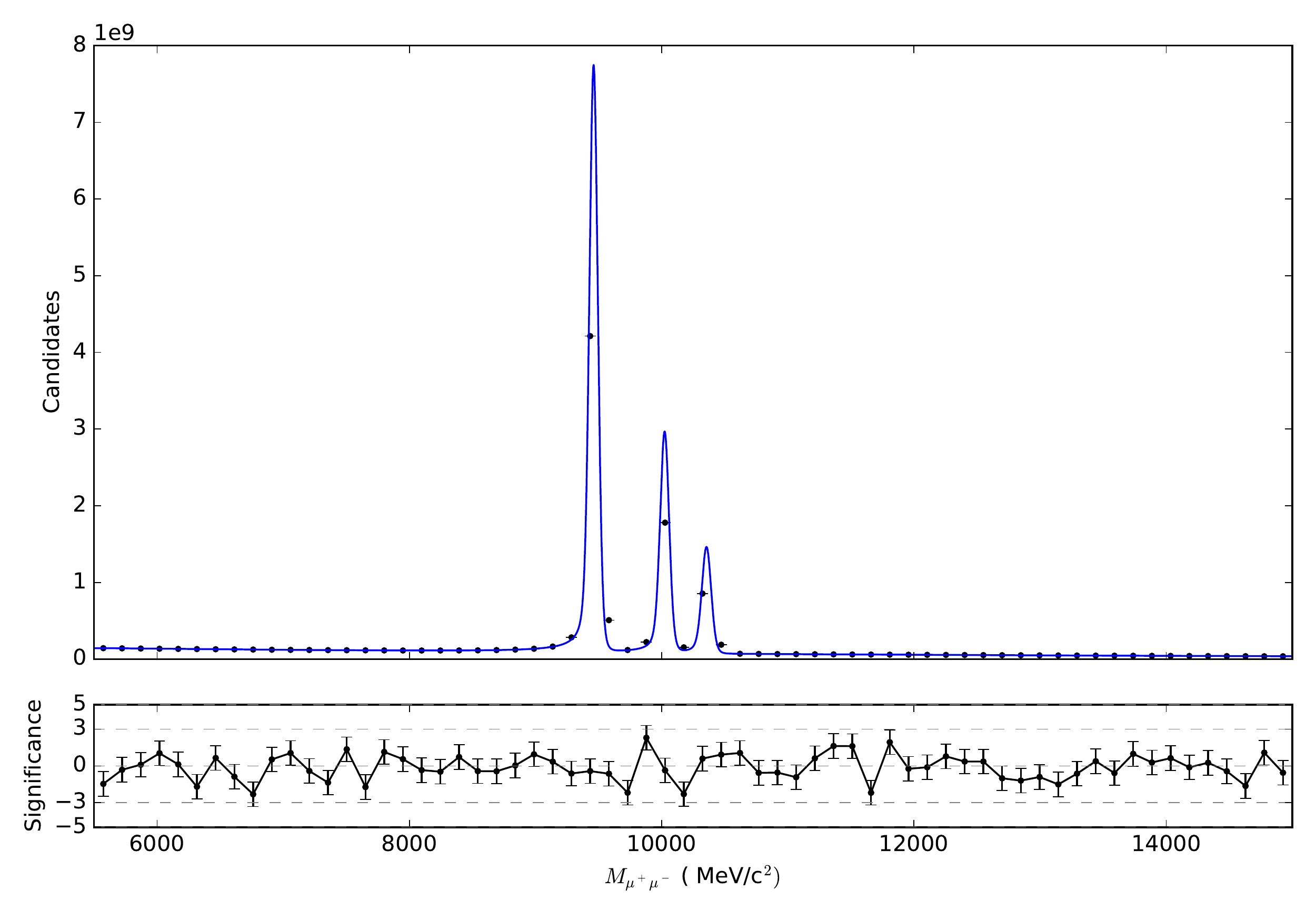}}\\
    \subfloat[$\text{Bin}_{p_T} = 1~~~~\text{Bin}_{\eta} = 0$]{
      \includegraphics[scale=0.25]{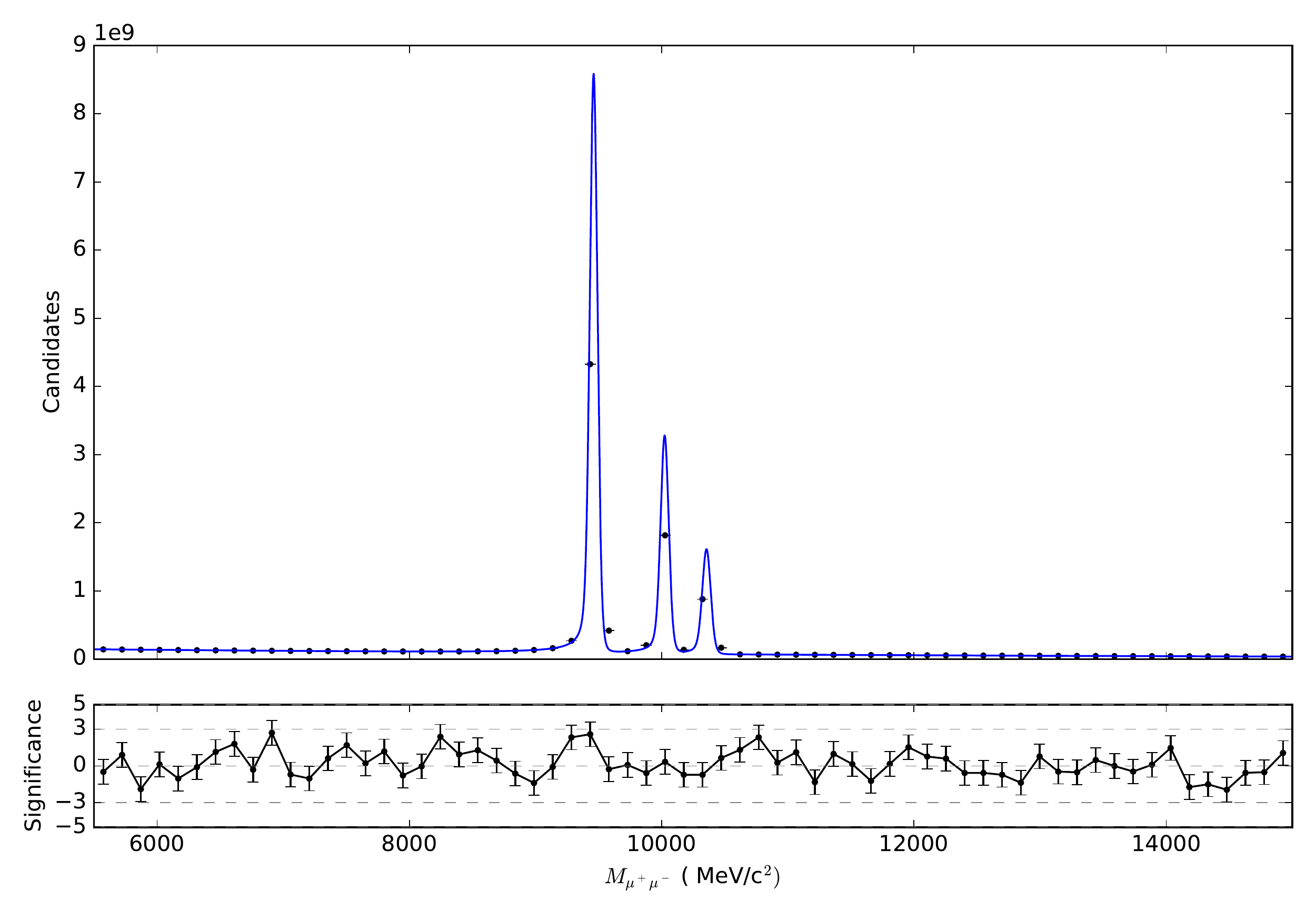}}
    \subfloat[$\text{Bin}_{p_T} = 1~~~~\text{Bin}_{\eta} = 1$]{
      \includegraphics[scale=0.25]{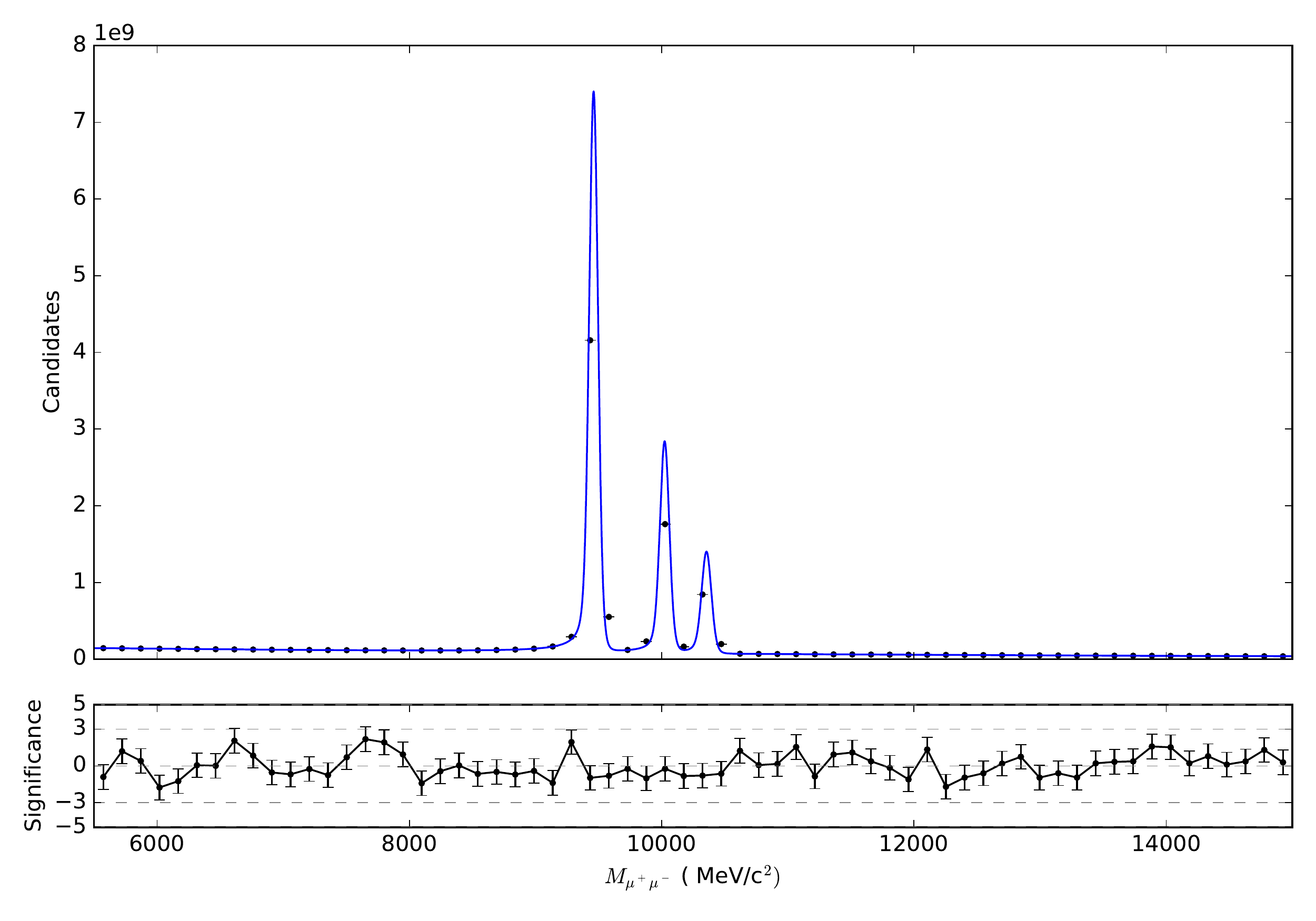}}
    \caption{
      Fits to the binned dataset of two opposite sign muons in the $\Upsilon$ range.
      The three peaks correspond to the three different states of the $\Upsilon$ resonance which contribute to this spectrum.
    }
    \label{fig:Example3}
  \end{center}
\end{figure}

File: {\verb ypsilons.py }\\
In this example, a binned fit is performed to the $\Upsilon$ resonances, on a decay into two muons of opposite sign.
The fit is performed simultaneously in bins of pseudorapidity, $\eta$, and transverse momentum, $p_T$.
The number of bins for these two variables is left to the user, and specified in two lists:
\begin{verbatim}
ptbins  = range(2)
etabins = range(2).
\end{verbatim}
Binning scheme for the data is set using
\begin{verbatim}
binning = gpuarray.to_gpu(np.float64(np.linspace(5.5e3, 15e3, 512*32)))
\end{verbatim}
where {\verb 5.5e3 } and {\verb 15e3 } are the bounds on the di-muon mass (in \mevcc) and {\verb 512*32 } defines a number of bins in terms of the grid used later.
Data is taken from a toy-MC sample, generated from a PDF as the one used in the fit.
The parameters to create the sample are specified in a dictionary
corresponding in this case to the intial values of those to do the fit:
\begin{verbatim}
genPars = {}                                                   
for p in Params:
  genPars[p.name] = p.val
manager.generateToy(genPars).
\end{verbatim}
For more information on how to generate toy-MC samples see Sect.~\ref{sec:toyMC}.
When the categories (one per bin) are defined:
\begin{verbatim}
Cat(cat_name, idata[1], getN = True)
\end{verbatim}
the input data for each category is set as binned specifying {\verb getN =True}.
The $\Upsilon$ fit parameters are fixed to those saved in {\verb params.py }. The free parameter limits are specified during the initialization of the variable \texttt{Params}:
\begin{verbatim}
Params = [
  Free('lm', 0.47, limits = (0., 0.9)),
  Free('lb', -2.0, limits = (-4, -0.1)),
  Free('a1', 1.26, limits = (0.5, 5.)),
  Free('n1', 1.28, limits = (0.5, 5.)),
  Parameter('n2', 1.),
  Parameter('a2', 200.),
  Free('beta', -0.001, limits = (-1.e-02, 0.)),
  Free('scale', 1.e-04, limits = (0, 0.005)),
  Parameter('A1_m', 9460.30),
  ]
\end{verbatim}
The mass model consists on a smeared Ipatia distribution, which inculdes a gaussian convolution for the multiple scattering ~\cite{Ipatia}.
The MS contribution is read from {\verb rampup_params.cpickle }.
As an output the mass plots for each $p_T$ and $\eta$ bin are shown and saved in pdf format (see Fig.~\ref{fig:Example3}).

\begin{table}[htb!]
\begin{center}
\caption{Timings of the Example 3 in different devices (see text for details).}
\label{tab:timings3}
\begin{tabular}{ccc}
\toprule
 Device & $10^{10}$ events\\
 & time [s]\\
\midrule
GeForce 980 GTX Ti &$\approx$ 50 \\
GeForce 1080 GTX &$\approx$ 50\\
Tesla M2090 & - \\
 \bottomrule
\end{tabular}
\end{center}
\end{table}

\section{Example 4: $\rm CL_s$ }
File: \texttt{cls\_beta\_01.py}\\
Limits setting and combinations of experimental results using the $\rm CL_s$ method (or \textit{modified frequentist approach}~\cite{CLs1,CLs2}) can greatly benefit from the use of GPUs, by parallelizing the computation of the toy experiments, that represent the most time-consuming part of the computation of the confidence levels. In this section a generic example is shown. 
\label{sec:exampleCLs}
This example consists on the computation of $\rm CL_b$,$\rm CL_{s+b}$ and $\rm CL_s$
in a hypothetical search experiment of a signal described by the Ipatia function~\cite{Ipatia} in the mass spectrum, on top of an exponential background. For this, a definition of the different parameters is done, such as the number of bins for the mass distribution (\texttt{NBINS = 10000}), as well as the the array of mass values that define the interval where the search is performed: \\
\texttt{masslow = np.arange(5179.,5379., 200./NBINS, dtype = dtype)\newline
DM = masslow[1] - masslow[0] \newline
massbins = np.append(masslow, np.array(max(masslow)+DM,dtype = dtype)) \newline
mass\_center = masslow +0.5*DM},\\
where \texttt{masslow} is the array of low-edges, \texttt{DM} its step size, \texttt{massbins} is defined from \texttt{masslow} such that the last value of the array is also incorporated, and \texttt{mass\_center} is the corresponding array with the bin centers.
After this, the Ipatia function is loaded into the GPU, and the values of its parameters are specified: \\
\texttt{mod = SourceModule(file(os.environ["IPAPATH"] + "/cuda/psIpatia.cu","r").read())\newline
ipa = mod.get\_function("Ipatia"). \newline
}
In order to convert unbinned data to histograms, bins for the GPU have to be defined from \texttt{mass\_center}:\\
\texttt{bins\_gpu = gpuarray.to\_gpu(mass\_center)\newline
bins\_cx = bins\_gpu.astype(np.complex128)}\\
Note that there are two sets of bins: the ones of type \texttt{np.float64} are used for both the Ipatia and exponential PDF that will be used later for the background, while the bins of type \texttt{np.complex128} will be necessary when convolving the Ipatia with a Gaussian function.\\
The array corresponding to the signal distribution is created as an empty array with the size of \texttt{mass\_center}, filled via the Ipatia function and then convolved with a Gaussian function as follows:\\
\texttt{ipa(bins\_gpu, sigf\_i\_gpu, np.float64(mu),np.float64(sigma),np.float64(l), np.float64(beta), np.float64(a), np.float64(n), np.float64(a2),\newline np.float64(n2) , block = (1000,1,1), grid = (max(len(tmp0)/1000,1),1)) \newline
SM = GaussSmear1(bins\_cx,sigma0)  \newline
sigf\_gpu = SM.Smear(sigf\_i\_gpu)}.\\
As for the background distribution, an array of the same length as \texttt{sigf\_gpu} is filled following an exponential function, $\exp(-\rm kx)$, where k is set to be -0.001 and x is the mass value. Both signal and background distributions are normalized to 1: \\
\texttt{sigpdf\_gpu = sigf\_gpu/np.sum((sigf\_gpu).get())\newline
bkgpdf\_gpu = bkgf\_gpu/np.sum((bkgf\_gpu).get()). } \\
From these PDFs, the data that will be used for the $\rm CL_s$ computation is generated, setting the number of expected events for both signal (\texttt{ns}) and background (\texttt{nb}).\\
\texttt{def HistoPdf(ns =5000, nb = Nb): return np.float64(ns) * sigpdf\_gpu + np.float64(nb)*bkgpdf\_gpu  \newline
fake\_data = gpuarray.to\_gpu(np.uint32(tmp0)) \newline
generateBinned(fake\_data, HistoPdf(ns = 70)).\newline}
Finally, the confidence levels and $\rm CL_s$ are calculated as a function of the \textit{signal strength} or number of signal events. The number of toy experiments used can be modified, and for the sake of this example is set to the default value (10000 toys), good enough to get proper precision. \\
\texttt{CL = CLcalculator()\newline
CL.setData(fake\_data)\newline
CL.setExpected(HistoPdf)\newline
CL.CLs(70, toys = 10000)
}
\\

\begin{table}[htb!]
\begin{center}
\caption{Timings of the example 4 in different devices (See text for details).}
\label{tab:timingsCLs}
\begin{tabular}{cc}
 Device & time $(s)$ per event\\ 
\hline
GeForce 980 GTX Ti & 10.6 \\
GeForce 1080 GTX & 12.8 \\
Tesla M2090 & 41.7 \\
 \bottomrule
\end{tabular}
\end{center}
\end{table}

\section{Example 5: Generating ToyMC}
\label{sec:toyMC}

GPU devices can also be used to generate toyMC. For binned PDFs, it can be done via \texttt{generateBinned} as 
shown in the examples of \CLs and $\Ah\to\mu\mu$.
Generating unbinned datasets require somewhat more gymnastics, but can also be done without too much effort, as we will
discuss now.
CUDA random generators are imported via \texttt{curand.h} and \texttt{curand\_kernel.h}. It must be noted that
the latter cannot be included inside an \texttt{extern "C"}, which is what \texttt{SourceModule} automatically wraps 
the code in. Thus, we have to invoke \texttt{SourceModule} (or \texttt{cuRead})
with the keyword argument \texttt{no\_extern\_c = True}. Then we should place \texttt{extern "C"} manually inside
our CUDA file, after the inclusion of the generators and before the function definitions.

Once this is done, one can proceed and use \texttt{curand\_init}, \texttt{curand\_uniform} etc to generate random numbers,
and create toyMC. The generation can be done for instance by using the {\it accept-reject} method.

An example of this is shown in file:
\begin{center}
\texttt{examples/toyMC/fitphis.py}
\end{center}

The above is a modified version of the example presented in~\secref{sec:example2}, where we generate a random dataset for one of the categories of
the fit, and plot it on top of the \pdf.
The example can easily be expanded to generate datasets for all categories, as well as to include Poissonian random
fluctuations on the generated number of events. For this we suggest generating \texttt{Nevts} in the \texttt{Python} class with \texttt{scipy.random.poisson}
or \texttt{scipy.random.normal}, and modifying the grid and block sizes accordingly.
The timing of the example measured in different setups is shown in \tabref{tab:timingstoy1}. 

\begin{table}[htb!]
\begin{center}
\caption{Timings of Example 5 in different devices (see text for details).}
\label{tab:timingstoy1}
\begin{tabular}{ccc}
\toprule
 Device & 133 725 events & 6 769 840 events\\
 & \multicolumn{2}{c}{time [s]}\\
\midrule
GeForce 980 GTX Ti & 0.11 & 0.34\\
GeForce 1080 GTX & 0.13 & 0.30\\
Tesla M2090 & - & - \\
 \bottomrule
\end{tabular}
\end{center}
\end{table}

\section*{Combining with Symbolic programing}

When dealing with complicated models, symbolic programing can be very useful. In \texttt{Python} this can be done through \texttt{sympy}. In that case, being able to convert a symbolic expression to a CUDA piece of code greatly simplifies the development of GPU fits. For instance, the angular PDFs and the corresponding integrals in the \BsJpsiKK example were obtained this way.
Another example of this is
\begin{center}
\texttt{examples/urania/faddeeva.py},
\end{center}
where the limit expression of Faddeva's functions at large $t/\sigma$ are obtained, and converted to \texttt{\_\_device\_\_} CUDA functions.

A more complicated ({\it and, sadly, cryptic}) example \footnote{Note: this was addapted to the sympy version quoted in~\secref{sec:others}, but the outcome for that version is has not yet been tested.} can be found at:
\begin{center}
\texttt{examples/urania/JpsiPhiTimeFit\_forRooFit.py}.
\end{center}
\noindent where the \BsJpsiPhi time-dependent angular distribution is printed to \texttt{LaTex} and \texttt{.pdf} files as well as to a \texttt{RooFit} fitting class. By enabling the flag \texttt{USE\_MATHEMATICA} one can get the integral over time
          from \texttt{Mathematica} instead of \texttt{sympy}, as the latter is sometimes worse at simplification. The \texttt{Python} list \texttt{Spins} lists the spins of the
          $hh$ resonances one wants to generate, i.e, \texttt{[1]} for \pwave, \texttt{[0,1]} for \swave and \pwave,  \texttt{[0,1,2]} to also include \dwave, etc\footnote{ A priori the list of waves is unlimited. But of course as the expression becomes more and more complicated, the symbolic manipulation may get stuck}...
          One can similarly make an script to create CUDA code. The following examples:
 \begin{center}
   \texttt{examples/urania/JpsiPhiTimeFit\_forCuda.py}\\
   \texttt{examples/urania/JpsiPhiTimeFit\_forCuda\_Bbar.py}\\
 \end{center}

 \noindent create CUDA files with the function, the integrals and the derivatives.

\section{FAQ}
 \label{sec:faq}
 \begin{itemize}
 \item {\bf Can I add prints/std::outs to my cuda stuff?}: Yes, \texttt{\#include $<stdio.h>$} and you could use \texttt{printf} at least on the \texttt{\_\_global\_\_} functions. You can also pass a
   \texttt{gpuarray} to your program and store on it intermediate numbers that you want to check, it will be visible from \texttt{Python}.
 \item 
   
 \end{itemize}
 
\clearpage
\newpage

\appendix
\cleardoublepage
\newpage
\section{\texttt{ParamBox} class}
\label{sec:ParamBox}

The class \texttt{ParamBox} is defined in \texttt{ModelBricks.py}. To initizalize an instance, one needs
to provide a list of fit parameters and (optionally) a list of categories (class \texttt{Cat}). The following
methods are contained in \texttt{ParamBox}:
\begin{itemize}
 \item \texttt{freeThese(pars)}: Frees each parameter in the provided list
 \item \texttt{lock\_to\_init(pars)}: Fixes each parameter in the provided list
 \item \texttt{createFit()}: Initializes a \texttt{Minuit} fit providing it with a list of parameters
 \item \texttt{createMultinest(savedir,reset = False,**kwargs)}: Runs a \texttt{Multinest} scan with the fit result as an input and saves 
 the output in a directory 
 \item \texttt{readMultinest(savedir)}: Reads an existing \texttt{Multinest} output and prints its result
 \item \texttt{fitSummary()}: Creates an instance of the class \texttt{FitSummary} (defined in the same file), 
 which contains the fit result information, such as the parameter values, parameter errors, fit matrix etc.
 \item \texttt{saveFitSummary(name)}: Saves the fit result information into a \texttt{cPickle} file with the name provided
\end{itemize}

\cleardoublepage
\newpage
\section{\texttt{Parameter} class}
\label{sec:Parameter}

The class \texttt{Parameter}, defined in \texttt{ModelBricks.py}, handles the information of each parameter in the fit. It is initialized with a name and, optionally, an initial value, range limits, a step size, a flag indicating whether it is constant (default), and a type. The following
methods are contained in \texttt{Parameter}:
\begin{itemize}
\item \texttt{setVal(var)}: Sets the initial value
\item \texttt{autoStepSize()}: Sets the step size to one tenth of the parameter range (determined by the limits)
\item \texttt{setLimits(m,M, constant = False)}: Sets the lower and upper limits and, optionally, changes the constant flag
\item \texttt{getSettings()}:  Returns a dictionary with the available information of the parameter (value, limits, error and whether it is constant)
\end{itemize}

The class \texttt{Free} is completely equivalent to \texttt{Parameter} but with the option \texttt{constant=False} set by default.
\cleardoublepage
\newpage
\section{\texttt{Cat} class}
\label{sec:Cat}

The information in each category of a simultaneous fit is handled by the class \texttt{Cat}, defined in \texttt{ModelBricks.py}. An instance of this class is initialized with a name for the category. Additionally, a dataset (GPU array, numpy ndarray, list, cPickle file or string) can be provided. If \texttt{getN=True} is added as extra argument, the class will determine the size of the given sample. Alternatively, the user can specify the size by setting \texttt{N} to the desired value in the arguments.

When an instance of the class \texttt{Cat} is initialized, the following two methods are called (if possible):
\begin{itemize}
\item \texttt{setData(ary, getN = False)}: Creates a GPU array to store the provided dataset
\item \texttt{bookProbs(N)}: Books a GPU array with the size of the data for the likelihood probability values at each point
\end{itemize}
\cleardoublepage
\newpage
\section{Hardware}
\label{sec:Hardware}

Here, we give some hardware specifications for the graphics cards used to test the examples in this documentation.

\begin{table}[htb!]
\begin{center}
\caption{Specifications of the graphics cards used for testing the examples.}
\label{tab:hardware}
\begin{tabular}{llll}
 \toprule
  & GeForce 980 GTX Ti & GeForce 1080 GTX & Tesla M2090\\
 \midrule
 CUDA cores & 2816 & 2560 &  512\\
 Single precision (GFLOPS) & 5632 & 8228 (8873) & 1331.2\\
 Double precision (GFLOPS) & 176 & 257 (277) & 665.6 \\
 Memory size (GiB) & 6.1 & 8 & 6.1 \\
 Memory Bandwidth (GB/s) & 336.5 & 320 & 177.6\\
 \bottomrule
\end{tabular}
\end{center}
\end{table}

\cleardoublepage
\newpage

\cleardoublepage
\newpage

\cleardoublepage
\newpage
\bibliographystyle{JHEP}
\bibliography{main}
\clearpage

\end{document}